\documentclass[10pt,english,draftclsnofoot,english,journal,onecolumn,12pt]{IEEEtran}
\usepackage[T1]{fontenc}
\usepackage[active]{srcltx}
\usepackage{color}
\usepackage{graphicx}

\makeatletter

\providecommand{\tabularnewline}{\\}

\ifCLASSOPTIONcompsoc
\usepackage[caption=false,font=normalsize,labelfont=sf,textfont=sf]{subfig}
\else
\usepackage[caption=false,font=footnotesize]{subfig}
\fi

\usepackage{cite}
\usepackage{bm}
\usepackage{algorithmic}
\usepackage{algorithm}
\usepackage{graphicx}
\IEEEoverridecommandlockouts

\@ifundefined{showcaptionsetup}{}{%
 \PassOptionsToPackage{caption=false}{subfig}}
\usepackage{subfig}
\makeatother

\usepackage{babel}
\begin{document}

\title{\textcolor{black}{Integrated Communication and Navigation for Ultra-Dense
LEO Satellite Networks: Vision, Challenges and Solutions}}

\author{\textcolor{black}{Yu Wang, Hejia Luo, Ying Chen, Jun Wang, Rong Li,
and Bin Wang }%
\thanks{\textcolor{black}{Y. Wang, H. Luo, Y. Chen, J. Wang, R. Li, and B.
Wang are with the Hangzhou Research Center, Huawei Technologies Co.,
Ltd., Hangzhou 310052, China.}%
}}

\maketitle

\section*{\textcolor{black}{Abstract}}

\textcolor{black}{Next generation beyond 5G networks are expected
to provide both Terabits per second data rate communication services
and centimeter-level accuracy localization services in an efficient,
seamless and cost-effective manner. However, most of the current communication
and localization systems are separately designed, leading to an under-utilization
of radio resources and network performance degradation. In this paper,
we propose an integrated communication and navigation (ICAN) framework
to fully unleash the potential of ultra-dense LEO satellite networks
for optimal provisioning of differentiated services. The specific
benefits, feasibility analysis and challenges for ICAN enabled satellite
system are explicitly discussed. In particular, a novel beam hopping
based ICAN satellite system solution is devised to adaptively tune
the network beam layout for dual functional communication and positioning
purposes. Furthermore, a thorough experimental platform is built following
the Third Generation Partnership Project (3GPP) defined non-terrestrial
network simulation parameters to validate the performance gain of
the ICAN satellite system. }
\begin{IEEEkeywords}
Integrated communication and navigation, multi-beam LEO satellite,
ultra-dense networks, positioning accuracy, beam hopping.
\end{IEEEkeywords}

\section{\textcolor{black}{Introduction\label{sec:introduction} }}

\textcolor{black}{ Accommodating ubiquitous connectivity and high
data rate communication services is one of the main goals for communication
networks. To complement this goal, satellite communication has gained
a renewed upsurge in the New Space era \cite{key-1} due to its ability
to provide global wireless coverage and continuous service guarantee
especially in scenarios not optimally supported by terrestrial infrastructures.
In particular, a new work item has recently been initialized by the
Third Generation Partnership Project (3GPP) to study a set of necessary
adaptations enabling the operation of 5G New Radio (NR) protocol in
non-terrestrial network (NTN) with a first priority on satellite access
\cite{key-2}. In the NTN context, compared with conventional geostationary
earth orbit (GEO) and medium earth orbit (MEO),  low earth orbit (LEO)
based satellite networks stand out as a promising solution considering
the lower propagation delay, power consumption and launch cost. As
such, numerous companies have announced ambitious plans to provide
broadband Internet access over the globe by deploying LEO satellite
mega-constellations, e.g., Oneweb, Kuiper and Starlink \cite{key-3}.}

\textcolor{black}{ In parallel to legacy communication services,
the rapid proliferation of location-based services, e.g., smart transportation,
augmented reality and autonomous driving, brings great value-added
opportunities to communication networks. In this regard, it is much
anticipated to provide high precision positioning and navigation}%
\footnote{\textcolor{black}{Navigation can be considered as a set of continuous
and dynamic positioning/localization procedures. In this paper, we
use navigation and positioning/localization interchangeably when no
ambiguity occurs. }%
}\textcolor{black}{{} information through communication networks \cite{key-4}.
As specified by NR Release-16, a positioning accuracy of 3-meter within
1 second end to end latency should be achieved for commercial use
cases \cite{key-5}, and subsequent releases are expected to further
targeting sub-meter accuracy and millisecond level lower latency.
To circumvent this issue, the vision of integrated communication and
navigation (ICAN) is proposed to fully reap the benefits of next generation
wireless networks \cite{key-6,key-7}. With ICAN, it becomes possible
to exploit co-design and optimization of communication and localization
aspects for best utilization of shared network infrastructures and
radio resources. }

\textcolor{black}{Specifically, in ultra-dense LEO satellite networks,
the ICAN paradigm exhibits some distinct advantages, such as navigation
improvement with stronger signal strength and better network geometry,
and user equipment (UE) self-localization enhanced network access
and mobility management. To the authors' best knowledge, this is the
first in-depth study on applying ICAN concept in the context of LEO
satellite mega-constellations. Particularly, we present a through
analysis on the benefits, feasibility and challenges for ICAN enabled
ultra-dense satellite networks. A novel beam hopping (BH) based solution
is then proposed to facilitate flexible ICAN satellite network operation,
such that the satellite system can adaptively tune its physical beams
for dual-functional communication and positioning purposes. The associated
measurement signal structure, physical layer control procedures and
practical BH algorithm are elaborated for the integrated satellite
system. To validate the performance gain brought by ICAN satellite
system, we have conducted a set of experiments following 3GPP NTN
satellite simulation assumptions. The numerical results demonstrate
that the proposed BH solution can dramatically improve the positioning
accuracy from approximate 200-meter to 20-meter. Next, some future
research directions are investigated. Finally, we conclude our work.}

\section{ICAN Satellite Network Vision Overview }

\subsection{\textcolor{black}{ICAN Satellite Network: Concepts and Benefits}}

\begin{figure}
\centering\includegraphics[scale=0.75]{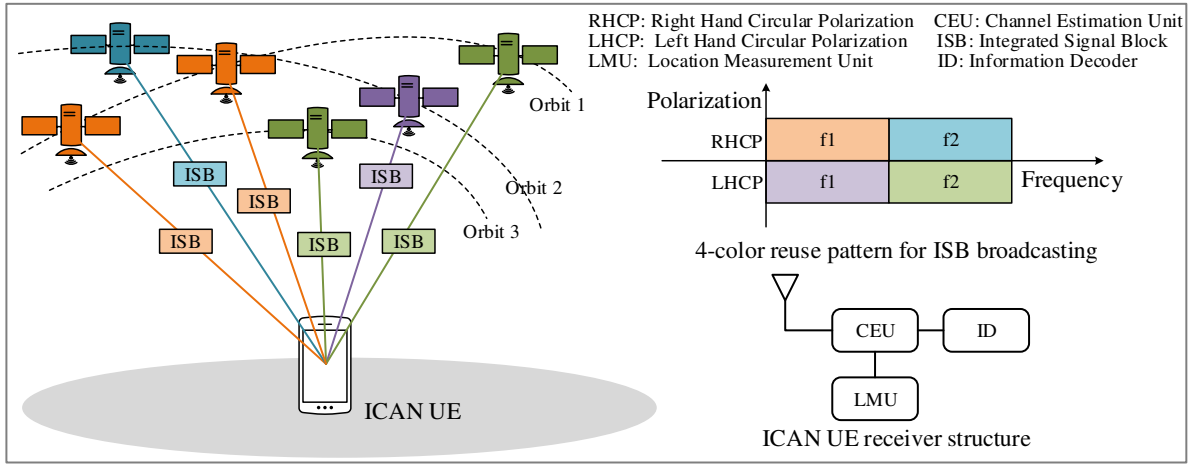}

\textcolor{black}{\caption{\textcolor{black}{An example of ICAN enabled satellite system with
4-color reuse and integrated receiver structure\cite{key-6}.\label{fig:ICANSat}}}
}
\end{figure}

\textcolor{black}{Traditionally, although localization systems can
reuse communication infrastructures like cellular networks cost-effectively
to some extent, the localization and communication systems are mostly
designed in separation. This inevitably results in a waste of precious
radio resources and network performance degradation. To this end,
the ICAN concept is proposed to support both communication and positioning
services with optimized network infrastructure design and radio resource
utilization for dual purposes. An illustrative example of ICAN enabled
satellite system is shown in Fig. \ref{fig:ICANSat}. An ICAN UE can
simultaneously receive a set of integrated signal blocks (ISBs) from
multiple satellites possibly operating at different orbits. ISBs are
utilized for both communication and localization purposes. Besides,
to alleviate inter-satellite interference, a color reuse pattern,
e.g., 4-color reuse, is employed to broadcasting ISBs originated from
different satellites/beams, wherein a color refers to a combination
of polarization and frequency. Last but not the least, an integrated
ICAN UE receiver structure can be customized to encompass channel
estimation, information decoder and location measurement \cite{key-6},
such that both services can be provisioned in a seamless manner.}

\textcolor{black}{Especially in the context of ultra-dense LEO satellite
networks, ICAN paradigm exhibits some distinct benefits as follows.}
\textcolor{black}{}

\textbf{\textcolor{black}{Location based communication optimization:
}}\textcolor{black}{Due to the prevalent frequency/polarization reuse
pattern, a UE can simultaneously measuring positioning reference signals
at different frequencies/polarizations from multiple satellites. Through
the range measurements and available broadcasting satellite ephemeris,
creditable self-localization can be attained by computing location
results at the UE side. The acquired location information can be exploited
to simplify network mobility management, improve Doppler compensation
and timing advance maintenance, and facilitate location enhanced network
access over the globe in the highly dynamic LEO satellite scenario.
As a consequence, network signaling overhead can be significantly
reduced.}

\textbf{\textcolor{black}{Communication for enhanced localization:}}\textcolor{black}{{}
Compared with traditional global navigation satellite system (GNSS)
operating at MEO and GEO, LEO satellite network offers several advantages
such as lower propagation delay, 300 to 2400 times stronger signals
and threefold improvement in satellite geometry\cite{key-8}. Through
exploitation of LEO satellite communication capabilities, e.g., satellite-air-ground
cooperative transmission, inter-satellite link augmented communication,
and satellite communication signals of opportunities, positioning/navigation
accuracy can be significantly improved and both multiple-satellite
searching complexity and time to first fix can be decreased to some
extent. }

\textcolor{black}{In summary, the adoption of ICAN can bring additional
potential to LEO satellite networks.}

\subsection{Feasibility Analysis }

One fundamental prerequisite towards realizing ICAN satellite system
is feasibility analysis. To this end, we highlight several conditions
and technical advances that accelerate the pace to a truly functional
integrated network.
\begin{itemize}
\item \textbf{\textcolor{black}{Satellite network densification:}}\textcolor{black}{{}
Driven by advanced manufacturing technology and cheaper launch costs,
the LEO satellite network scale has tremendously expanded from traditionally
tens of satellites to thousands or even tens of thousands of satellites
in the last few decades. For instance, Amazon has recently gained
the approval of its Kuiper constellation with 3236 satellites from
the U.S. Federal Communications Commission (FCC) \cite{key-9}. Meanwhile,
SpaceX tends to build its magnificent Starlink constellation composed
of nearly 12000 satellites, with more than 500 satellites already
deployed on orbit by June 2020 \cite{key-10}. As a matter of fact,
a UE in a dense satellite network generally can have tens or even
hundreds of satellites in view simultaneously. This greatly boosts
the communication and navigation capability through multi-satellite
cooperation technique.}
\item \textbf{\textcolor{black}{Satellite communication characteristics:}}\textcolor{black}{{}
Unlike the aggressive full frequency reuse in cellular networks, satellite
communication typically adopts a frequency/polarization reuse pattern
with reuse factor larger than one to mitigate inter-beam interference.
This fact is consistent with the localization requirement, as a UE
can receive multiple signals at the same time by monitoring different
frequencies/polarizations. On the other hand, different from terrestrial
networks where the location of base stations is confidential and not
informed to UEs, satellite ephemeris are commonly available to UEs
in coverage. Therefore, in the satellite scenario, it is feasible
and convenient to compute location at the UE side without incurring
much complexity and signaling overhead.}
\item \textbf{\textcolor{black}{Flexible onboard payload:}}\textcolor{black}{{}
Owing to the development of flexible satellite payload technique,
satellite communication has evolved from previous bent-pipe/transparent
architecture to current regenerative one with powerful onboard processing
functions, e.g., beam steering and resource orchestration ability
\cite{key-11}. As a consequence, multiple functions can now be implemented
within the satellite itself rather than the ground stations. For example,
an Iridium-Next satellite is equipped with communication payload,
navigation enhancement payload and earth observation sensors. On the
basis, intelligent payload operation for multifarious services can
be performed.}
\end{itemize}

\subsection{Challenges }

\textcolor{black}{Despite the above advantages, it is technically
challenging to design an ICAN satellite system.}
\begin{itemize}
\item \textbf{Flexible system architecture:} Currently, a communication
system is designed mainly for communication purpose with proprietary
payload and protocols. As an integrated network, the network is expected
to intelligently reconfigure its architecture, e.g., network geometry,
reference signal structure and control procedure, such that differentiated
services can be optimally supported. Unfortunately, the dynamic network
topology variation and constrained multi-dimensional heterogeneous
resources induce difficulties in system architecture design.
\item \textbf{Quality of Service (QoS) provisioning:} Communication services
and positioning services have distinct QoS requirements. Regarding
communication services, it is sufficient to have good coverage and
signal quality purely from the serving satellite. While for positioning
services, acceptable signal quality and network geometry of multiple
satellites are essential. The aforementioned two goals are \textcolor{black}{contradictory
in general, because a good signal quality for the serving satellite
normally means low interference as well as signal strength from neighboring
satellites.}
\item \textbf{Beam management:} Next generation satellites are expected
to carry several tens of narrow spot beams. Those beams can be manipulated
over complex dimensions, e.g., space, time, power and frequency domain.
Multi-dimensional resource joint optimization should be performed
for beam management to deal with the extremely non-uniform traffic
distribution. To make matter worse, with the fast mobility of satellites,
i.e., approximate 7.5 km/s for a typical LEO satellite, beams should
be shut on and off from time to time for interference reduction. This
leads to extra non-negligible complexity in satellite beam management.
\end{itemize}

\section{\textcolor{black}{Beam Hopping based ICAN Satellite System }}

\textcolor{black}{Herein, we mainly elaborate on the novel BH based
ICAN satellite system framework, which includes the system workflow,
detailed BH algorithm design and performance evaluation. }

\subsection{\textcolor{black}{System Workflow}}

\textcolor{black}{First, both the downlink measurement signal structure
and physical layer control procedures are designed to facilitate the
ICAN satellite system operation. }

\subsubsection{New Measurement Signal Structure}

\begin{figure}
\centering\subfloat[]{\includegraphics[scale=1.3]{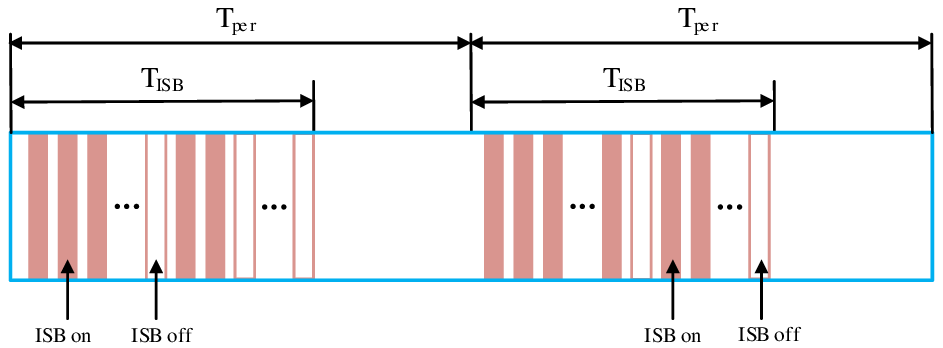}}

\subfloat[]{\includegraphics[scale=1.3]{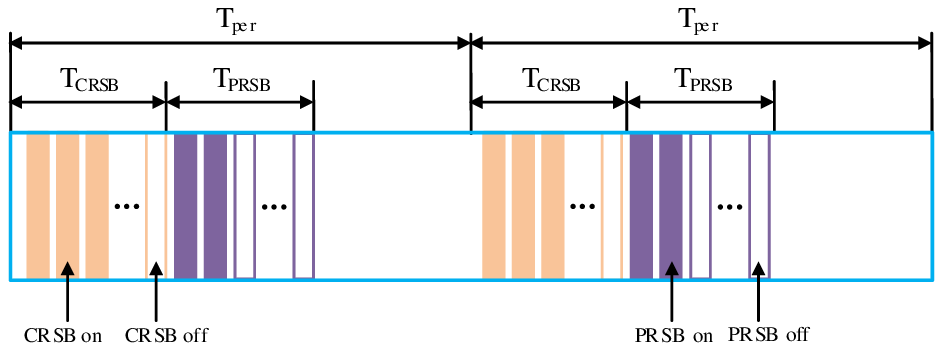}}

\textcolor{black}{\caption{\textcolor{black}{Measurement signal structures for ICAN satellite
system: (a) ISB with same CRSB and PRSB design case; (b) ISB with
separate CRSB and PRSB design case.\label{fig:Measurement-signal-structure}}}
}
\end{figure}

In NR, the downlink measurement signals for UEs during initial access
and idle mode are termed as synchronization signal blocks (SSBs).
While in the ICAN enabled NTN environment, a new measurement signal
structure is required to enable both cell search and UE passive localization
process. Specifically, the ISBs consist of two logical parts, namely\textbf{\textcolor{black}{{}
}}\textcolor{black}{communication reference signal blocks (CRSBs)
and} \textcolor{black}{positioning reference signal blocks (PRSBs)}.
We use the term logical because the actual realization of CRSBs and
PRSBs can be the same in uniform broadcasting signal design case or
different in separate broadcasting signal design case. An example
measurement signal structure for the above two cases are depicted
in Fig. \ref{fig:Measurement-signal-structure}(a) and Fig. \ref{fig:Measurement-signal-structure}(b),
respectively.\textcolor{black}{{} Hereafter, the separate design case
is taken as an example to show the concrete functions of CRSBs and
PRSBs.}

\textbf{\textcolor{black}{CRSBs:}}\textcolor{black}{{} The CRSBs are
used for UEs to acquire time and frequency synchronization with a
cell and to detect its physical layer cell ID. The CRSBs last for
a time duration of $T_{CRSB}$ within a periodicity of $T_{per}$.
To some extent, one can view CRSBs as conventional SSBs, except that
CRSBs have a more general meaning and can span over frequency/time/space
domain. Notably, a UE can acquire the PRSB configurations for multiple
neighboring satellites by extracting the system information (SI) contained
in the CRSBs of the accessed cell.}

\textbf{\textcolor{black}{PRSBs:}}\textcolor{black}{{} The PRSBs are
dedicated for ranging purpose, and span for a time duration of $T_{PRSB}$
within a periodicity of $T_{per}$. The pseudo-random quadrature phase
shift keying (QPSK) sequence with diagonal pattern defined in Release-9
is an example PRSB design. A UE can receive PRSBs from multiple satellites
on demand and perform related measurements, e.g., time of arrival
(ToA), frequency of arrival (FoA), and angle of arrival (AoA), to
calculate self location. Note that both CRSBs and PRSBs can be turned
on and off for the aim of interference reduction and service provisioning.
Furthermore, the time duration of $T_{CRSB}$ and $T_{PRSB}$ can
be tuned dynamically to strike a good balance between communication
and localization performance.}

\subsubsection{Physical Layer Control Procedures}

\begin{figure}
\centering\includegraphics[scale=0.9]{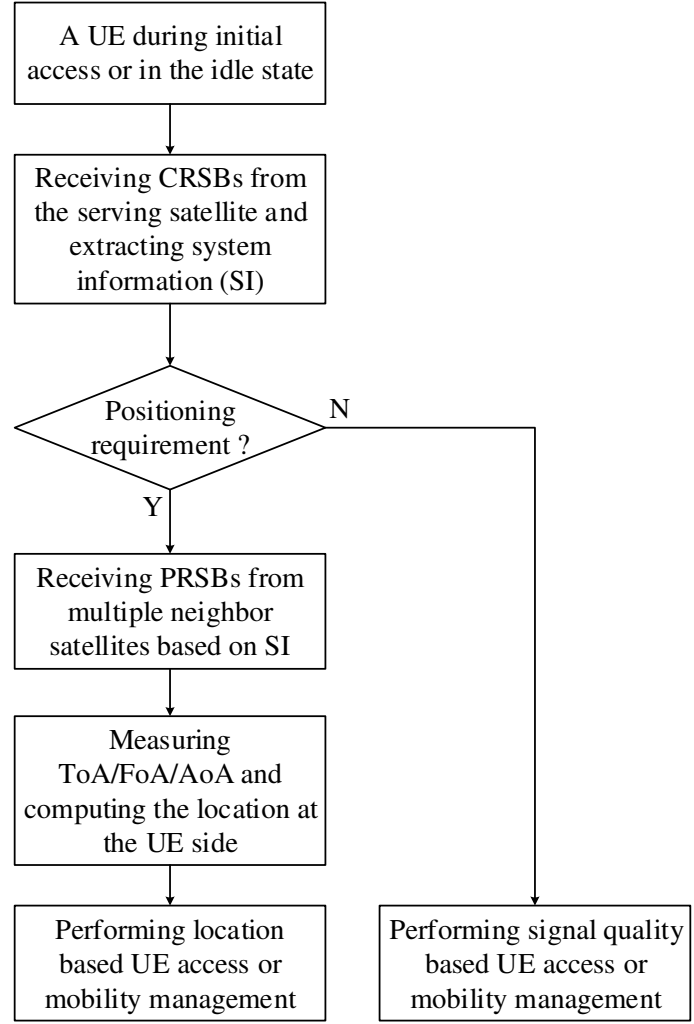}\textcolor{black}{\caption{Physical control procedure for ICAN satellite system.\label{fig:Physical-control-procedure}}
}

\end{figure}

\textcolor{black}{The physical layer control procedure is tailored
for ICAN satellite system. A concrete flow diagram of the control
procedure is given in Fig. \ref{fig:Physical-control-procedure}.
To be specific, a UE during initial access or in the idle mode should
first search for the CRSBs from its serving satellite. Based on the
received signal quality and CRSB configuration information, the UE
can then perform random access or cell reselection procedure. Note
that the aforementioned procedures are completed purely based on signal
quality measurements, e.g., reference signal receiving power (RSRP).
If there is positioning requirement for mobility management or cell
access, the UE should then measure PRSBs from multiple neighboring
satellites besides the serving satellite. The configuration information
of the PRSBs can be obtained from the serving satellite's SI. Through
adequate number of range measurements, e.g., ToA/FoA/AoA, the UE is
able to calculate its own location and then utilize the location information
for various usage, e.g., location enhanced network access. }

\subsection{Beam Hopping Scheme }

\textcolor{black}{Due to the payload weight and volume limitations,
it is essential to employ only a small subset of transmitters/beams
for serving extensive satellite coverage area. As an appealing solution,
BH has been proposed accordingly. Assuming that a satellite intends
to achieve a coverage of $N$ beams by $K$ transmitters/beams with
$K<N$. For this aim, at each time stamp, a maximum number of $K$
beams are assigned to illuminate a portion of the whole satellite
coverage area, and time-division multiplexing approach is implemented
to manipulate the set of $K$ beams into different portions within
the coverage area of $N$ beams. Through flexible beam allocation,
full satellite coverage service can be eventually obtained. }
\begin{figure}[t]
\centering\subfloat[]{\includegraphics[scale=0.55]{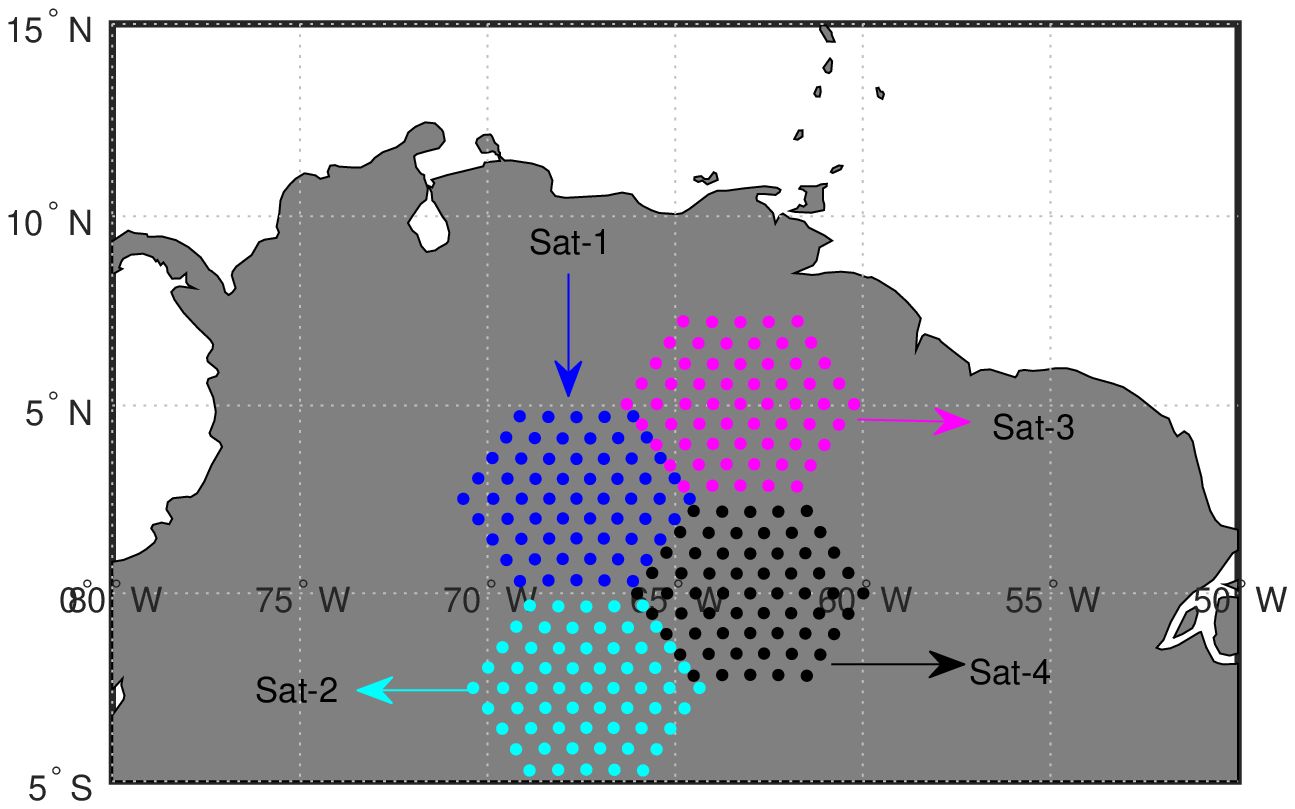}}\subfloat[]{\includegraphics[scale=0.55]{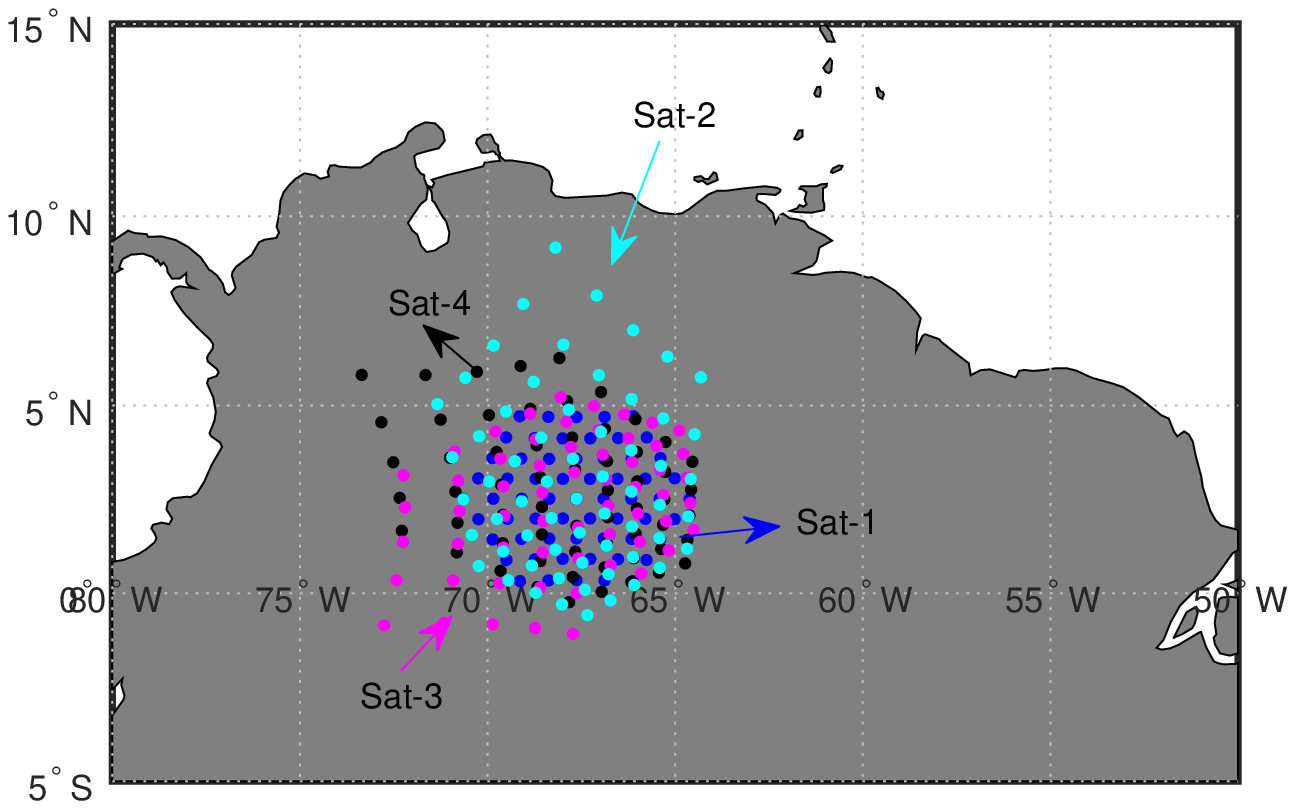} }\textcolor{black}{\caption{Beam layout for an example ICAN satellite system: (a) communication
beams; (b) positioning beams. Dots of the same color correspond to
the sets of beams in a satellite, with each dot representing a certain
beam center mapping on the geodetic plane.\label{fig:Beam-layout} }
}
\end{figure}

\textcolor{black}{Herein, we exploit the BH mechanism to efficiently
broadcast ISBs, i.e., CRSBs and PRSBs, in ICAN enabled satellite systems.
More specifically, CRSBs are transmitted using communication beams,
while PRSBs are transmitted by positioning beams. To support the beam
sharing over a common physical transceiver, BH is applied to adaptively
tune the physical beams between communication beams and positioning
beams in a time-division multiplexing manner. Besides, an efficient
UV plane based BH algorithm (UVBHA) is devised for beam transformation,
where UV plane is defined as the perpendicular plane to the satellite-earth
line on the orbital plane \cite{key-2}. In UVBHA, a hexagonal beam
layout is defined on the UV plane with UV coordinate of the nadir
of the reference satellite setting to (0,0) for communication beams.
For positioning beams, there are two different configuration situations.
Firstly, to perform localization in its own service area, the satellite
can simply reuse the communication beams as positioning beams. Secondly,
to assist localization for a neighboring satellite, the satellite
needs to translate the beams centered at (0,0) to the center of the
neighbor satellite's nadir in the UV plane denoted by $(u,v)$. We
can derive $u=sin\theta cos\varphi$ and $v=sin\theta sin\varphi$,
where $\theta$ and $\varphi$ represent beam bore-sight steering
angle and azimuth, respectively.}

\textcolor{black}{An example beam layout of 4 satellites for the proposed
UVBHA in geodetic plane is shown in Fig. \ref{fig:Beam-layout}. As
can be seen, each satellite is equipped with 61 beams. The set of
communication beams are pointed at the origin, i.e., (0,0) in the
UV plane, of the reference satellite. In addition, Sat-1 in Fig. \ref{fig:Beam-layout}(b)
still uses the set of communication beams as positioning beams. For
neighboring satellites, i.e., Sat-2, Sat-3 and Sat-4, the set of beams
are redirected to the nadir of Sat-1 to enable UE localization in
its service area. Consequently, with UVBHA, the physical beams can
be shared for dual purposes, i.e., communication and positioning beams.
There are several aspects to be noted in BH algorithm design. First,
the beam layout for positioning beams are generally not hexagonal
as beams deviates from the nadir. Second, the coverage of positioning
beams after BH is larger than that of communication beams, and thus
further optimization, e.g., turn off some edge beams, can be investigated.}
\begin{table}[h]
\textcolor{black}{\caption{\textcolor{black}{Key simulation parameters\label{tab:Key-simulation-paramters.}. }}
~~~~~~~~~~~~~~~~~~~~~~~~~~~~~~~~~~}%
\begin{tabular}{|l|l|}
\hline
Parameters  & \textcolor{black}{Values}\tabularnewline
\hline
\textcolor{black}{The number of orbit} & \textcolor{black}{40}\tabularnewline
\hline
The number of satellite per orbit & 60\tabularnewline
\hline
Orbit inclination & \textcolor{black}{87.5}\tabularnewline
\hline
\textcolor{black}{Orbit height} & 1200 km\tabularnewline
\hline
The number of beam per satellite  & 61\tabularnewline
\hline
Carrier frequency & 2 GHz\tabularnewline
\hline
Satellite EIRP density & 40 dBW/MHz\tabularnewline
\hline
System bandwidth & 30 MHz\tabularnewline
\hline
Frequency reuse factor & 3\tabularnewline
\hline
Equivalent satellite antenna aperture & 0.5 m\tabularnewline
\hline
Antenna pattern & Bessel function \cite{key-2}\tabularnewline
\hline
Channel Model & Clear Sky with line of sight\tabularnewline
\hline
UE type  & Handheld\cite{key-2}\tabularnewline
\hline
CRSBs/PRSBs & SSBs\tabularnewline
\hline
Subcarrier Spacing & 15 KHz\tabularnewline
\hline
Positioning method  & TDOA\tabularnewline
\hline
Number of satellites for positioning & 6, 8\tabularnewline
\hline
\end{tabular}
\end{table}

\subsection{\textcolor{black}{Performance Evaluation}}

\textcolor{black}{}
\begin{figure}
\centering\includegraphics[scale=0.8]{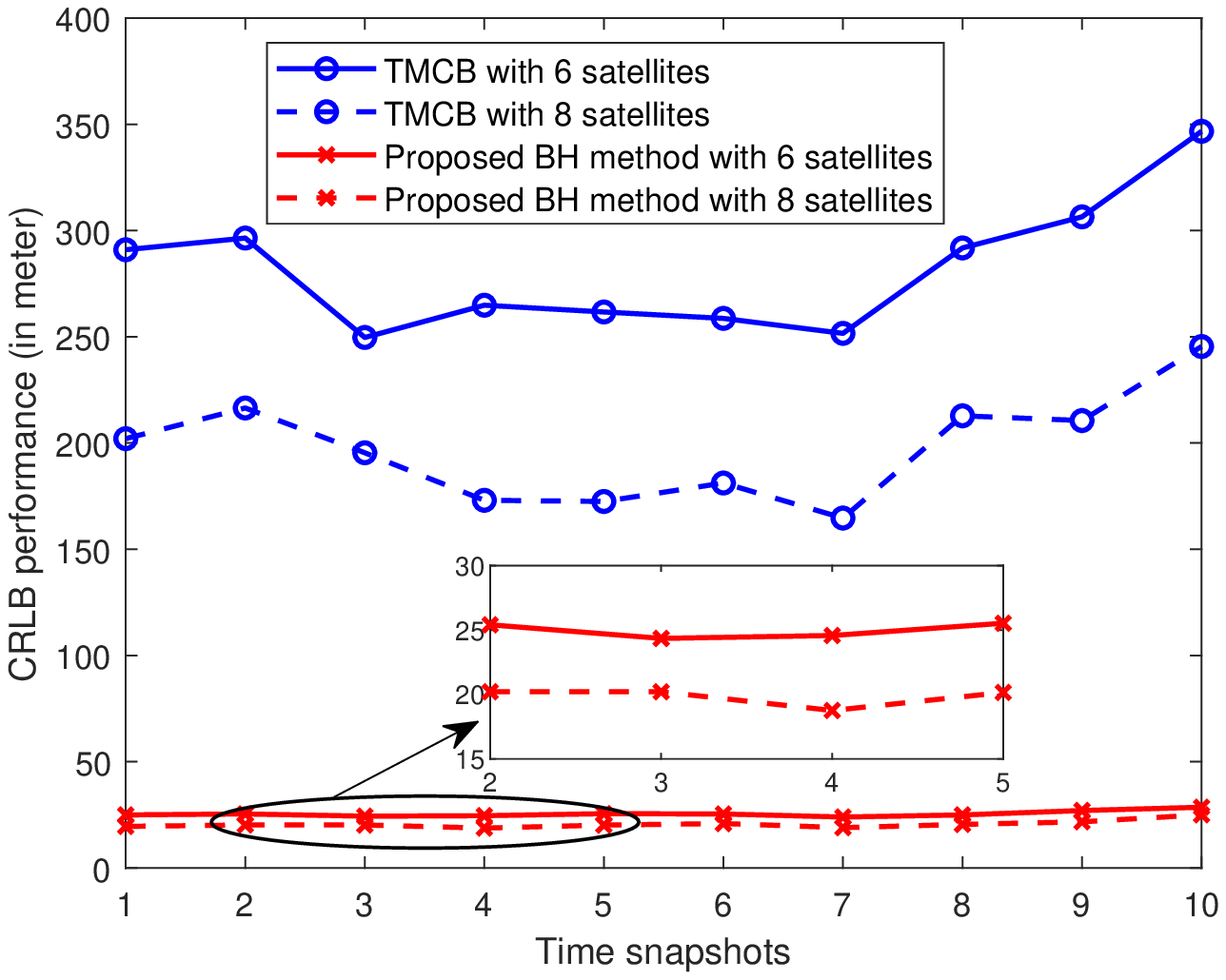}\textcolor{black}{\caption{CRLB results for different schemes in the example ICAN satellite network.\label{fig:CRLB-results.}}
}
\end{figure}
In this part, to validate the performance gain brought by ICAN satellite
system, a thorough simulation platform is built based on 3GPP NTN
simulation assumptions and parameters. The set of key parameters used
for simulation are summarized in Table \ref{tab:Key-simulation-paramters.}.
Particularly, the satellite network comprises a total of 2400 satellites,
with an orbit height of 1200 km and inclination of 87.5 degree. For
dynamic simulation, the orbit period is divided into 100 snapshots
of equal time duration. A total of 500 stationary UEs are randomly
deployed in the target area with longitude and latitude setting to
{[}-70,-60{]} and {[}-5,5{]}, respectively. The time differential
of arrival (TDOA) based positioning algorithms adopted by 3GPP technical
specifications are taken as the benchmark and the corresponding Cramer-Rao
lower bound (CRLB) for positioning performance is theoretically derived%
\footnote{Since the communication beams in conventional system and ICAN satellite
system are identical, there is no difference in communication related
performance, e.g., signal quality. We thus only consider positioning
related performance comparison in this paper.%
}. \textcolor{black}{The CRLB results for different schemes in the
example network are given in Fig. \ref{fig:CRLB-results.}. It can
be observed that the proposed BH method significantly outperforms
traditional method using communication beams (TMCB) in terms of CRLB.
This is because in TMCB, although the received signal quality from
the serving satellite is favorable, the signal quality from neighboring
satellites is very poor due to the long distance between UE and beam
center location. Nonetheless, in the proposed BH method, neighboring
satellite beam center is directed to cover UEs of interest, and thus
multiple signals with good quality can be measured to facilitate high
accuracy positioning. Besides, as the number of positioning satellites
increases from 6 to 8, the CRLBs for both algorithms improve as well.
This phenomena can be expected, because more signals and better network
geometry are obtained by exploring satellite diversity. Furthermore,
both the TMCB and proposed BH method exhibit CRLB fluctuation as time
snapshots change. The variation tendency is quite complicated, due
to the intertwined influence by time varying inter-beam interference
and dynamic geometric dilution of precision (GDOP).}

\section{\textcolor{black}{Open Research Issues}}

\textcolor{black}{In this section, we identify several open research
directions related to the ICAN system design.}

\subsection{\textcolor{black}{Intelligent Multi-Resource Allocation}}

With multi-dimensional resources, e.g., power, time, frequency and
beam, presented in the integrated network, intelligent resource allocation
scheme is highly required. However, due to the highly dynamic network
topology evolution, the multi-resource availability varies accordingly.
Besides, the non-uniform traffic distribution in both space-time domain
makes the optimal match between resource and differentiated services
challenging. To make matter worse, the use of massive satellites and
dense beams exacerbates the prevalent inter-beam interference problem
in satellite networks. This issue needs to be carefully tackled in
the resource allocation schemes. One viable solution is to apply the
promising machine learning technique, e.g., deep learning and reinforcement
learning, because more cognition and intelligence can be exploited
to improve network performance under complex and dynamic radio conditions
\cite{key-12,key-13}.

\subsection{Integrated Broadcasting Signal Design}

\textcolor{black}{To fully reap the benefits of the envisioned ICAN
satellite network, a fundamental issue lies in the ISB design for
dual communication and navigation usage. As for communication waveform
design, several key parameters including peak-to-average power ratio
(PAPR), out of band leakage, error performance and spectral efficiency
should be considered. With respect to navigation waveform design,
it is crucial to take autocorrelation/cross-correlation properties
and ranging resolution requirements into account. However, traditional
orthogonal frequency division multiplexing (OFDM) communication signal
is suboptimal in the satellite environment, due to the high PAPR,
large Doppler shift and long delay presented in ICAN satellite system.
Further investigation is highly demanded to solve the aforementioned
problem.}

\subsection{Theoretical Performance Analysis Framework }

\textcolor{black}{Theoretical performance analysis can provide important
guidance for ICAN satellite system design. However, it is a nontrivial
task to conduct theoretical performance analysis for ultra-dense LEO
satellite networks considering the following aspects. On the one hand,
the network is dynamic in terms of both channel connectivity and quality,
a tractable analysis framework dealing with those dynamics is troublesome.
Moreover, the complicated interaction between communication and navigation
should be captured in the performance analysis framework. On the other
hand, numerous imperfect stochastic effects should be accounted in
the ICAN satellite network for accurate analysis, such as large delay,
nonlinearity, orbit perturbation, atmospheric environment and so on.
A possible solution is to leverage both the stochastic geometry and
dynamic graph theory \cite{key-14,key-15}, such that both the dynamic
and stochastic effects are elegantly modeled. Overall, in-depth research
efforts are expected to tackle this issue.}

\subsection{Multi-Layer Heterogeneous Network Cooperation}

It is widely acknowledged that a single standalone network infrastructure
is not sufficient and economic to support diverse service requirements,
e.g., \textcolor{black}{Terabits per second data rate and centimeter-level
positioning accuracy. For instance, terrestrial cellular network is
generally not available in remote and disaster-striking areas, while
satellite/aerial communication experiences heavy attenuation in indoor
environments. In this regard, it is imperative to construct a multi-layer
heterogeneous network architecture integrating satellite, air, and
ground counterparts. However, due to the various characteristics and
capacity of the above three network segments, network cooperation
scheme with context-awareness becomes quite complex and remains to
be studied. }

\section{\textcolor{black}{Conclusion\label{sec:Conclusion}}}

\textcolor{black}{This article addressed the issue of exploiting ICAN
techniques for ultra-dense LEO satellite networks. The beam hopping
based framework encompassing reference signal structure, physical
layer control procedure, algorithm design, and performance evaluation
was comprehensively studied. Significant CRLB performance gains are
achieved through the integrated system. Since ICAN enabled satellite
system is still in its infancy, we expect that this study could open
up a new research direction for the design of next-generation LEO
satellite networks. }


\begin{thebibliography}{10}
\bibitem[1]{key-1}O. Kodheli \textit{et al.}, \textquotedblleft{}Satellite
Communications in the New Space Era: A Survey and Future Challenges,\textquotedblright{}
\textit{\textcolor{black}{arXiv preprint arXiv:2002.0}}\textit{8811},
pp. 1-45, Mar. 2020.

\bibitem[2]{key-2}3GPP TR 38.821 V16.0.0, \textquotedblleft{}Solutions
for NR to Support Non-Terrestrial Networks (NTN),\textquotedblright{}
Release 16, Dec. 2019.

\bibitem[3]{key-3} B. Di \textit{et al.}, \textquotedblleft{}Ultra-Dense
LEO: Integration of Satellite Access Networks into 5G and Beyond,\textquotedblright{}
\textit{IEEE Wireless Commun.}, vol. 26, no. 2, pp. 62-69, Apr. 2019.

\bibitem[4]{key-4}Jose A. del Peral-Rosado \textit{et al.}, ``Survey
of Cellular Mobile Radio Localization Methods: From 1G to 5G,'' \textit{IEEE
Commun. Surveys \& Tutorials}, vol.20, no.2, pp. 1124-1148, 2018.

\bibitem[5]{key-5}3GPP TR 38.855 V16.0.0, \textquotedblleft{}Study
on New Radio (NR) Positioning Support,\textquotedblright{} Release
16, Mar. 2019.

\bibitem[6]{key-6}Z. Xiao \textit{et al.}, \textquotedblleft{}An
Overview on Integrated Localization and Communication Towards 6G,\textquotedblright{}
\textit{arXiv preprint arXiv:2006.01535}, pp. 1-35, Jun. 2020.

\bibitem[7]{key-7}S. Jeong \textit{et al.}, \textquotedblleft{}Beamforming
Design for Joint Localization and Data Transmission in Distributed
Antenna System,\textquotedblright{} \textit{IEEE Trans. Veh. Technol.},
vol. 64, no. 1, pp. 62-76, Jan. 2015.

\bibitem[8]{key-8}T. Reid \textit{et al.}, \textquotedblleft{}Broadband
LEO Constellations for Navigation,\textquotedblright{} \textit{NAVIGATION,
Journal of the Institute of Navigation}, vol. 65, no. 2, pp. 205-220,
Feb. 2018.

\bibitem[9]{key-9}FCC, \textquotedblleft{}Application for Fixed Satellite
Service by Kuiper Systems LLC,\textquotedblright{} July 2020.

\bibitem[10]{key-10}FCC, \textquotedblleft{}SpaceX Non-Geostationary
Satellite System (Attachment A),\textquotedblright{} 2016.

\bibitem[11]{key-11}N. Zhang \textit{et al.}, \textquotedblleft{}Software
Defined Space-Air-Ground Integrated Vehicular Networks: Challenges
and Solutions,\textquotedblright{} \textit{IEEE Commun. Mag.}, vol.
55, no. 7, July 2017, pp. 101-109.

\bibitem[12]{key-12}P. V. R. Ferreira \textit{et al.}, \textquotedblleft{}Reinforcement
Learning for Satellite Communications: From LEO to Deep Space Operations,\textquotedblright{}
\textit{IEEE Commun. Mag.}, vol. 57, no. 5, pp. 70-75, May 2019.

\bibitem[13]{key-13}N. Kato \textit{et al.}, \textquotedblleft{}Optimizing
Space-Air-Ground Integrated Networks by Artificial Intelligence,\textquotedblright{}
\textit{IEEE Wireless Commun.}, vol. 26, no. 4, pp. 140-147, Aug.
2019.

\bibitem[14]{key-14}Y. Wang\textit{ et al}., \textquotedblleft{}Multi-Resource
Coordinate Scheduling for Earth Observation in Space Information Networks,\textquotedblright{}
\textit{IEEE J. Sel. Areas Commun.}, vol. 36, no. 2, pp. 268-279,
Feb. 2018.

\bibitem[15]{key-15}N. Okati \textit{et al}., \textquotedblleft{}Downlink
Coverage and Rate Analysis of Low Earth Orbit Satellite Constellations
Using Stochastic Geometry,\textquotedblright{} \textit{IEEE Trans.
Commun.}, vol. 68, no. 8, pp. 5120-5134, Aug. 2020.\end{thebibliography}
\end{document}